\documentclass[prl,twocolumn,showpacs,amsmath,amssymb,superscriptaddress]{revtex4}
\usepackage{graphicx}
\begin{document}
\preprint{}
\title {Possible re-entrant superconductivity in EuFe$_{2}$As$_{2}$ under pressure}
\author{C. F. Miclea}\email{miclea@cpfs.mpg.de}
\affiliation{Max Planck Institute for Chemical Physics of Solids, N\"{o}thnitzer Str. 40,
01187 Dresden, Germany}

\author{M. Nicklas} \email{nicklas@cpfs.mpg.de}
\affiliation{Max Planck Institute for Chemical Physics of Solids, N\"{o}thnitzer Str. 40,
01187 Dresden, Germany}

\author{H. S. Jeevan}\affiliation{I. Physik. Institut, Georg-August-Universit\"{a}t
G\"{o}ttingen, 37077 G\"{o}ttingen, Germany}

\author{D. Kasinathan}
\affiliation{Max Planck Institute for Chemical Physics of Solids,
N\"{o}thnitzer Str. 40, 01187 Dresden, Germany}

\author{Z. Hossain}
\affiliation{Department of Physics, Indian Institute of Technology, Kanpur 208016, India}

\author{H. Rosner}\affiliation{Max Planck Institute for Chemical Physics of Solids,
N\"{o}thnitzer Str. 40, 01187 Dresden, Germany}

\author{P. Gegenwart}
\affiliation{I. Physik. Institut, Georg-August-Universit\"{a}t
G\"{o}ttingen, 37077 G\"{o}ttingen, Germany}
\author{C. Geibel}
\affiliation{Max Planck Institute for Chemical Physics of Solids,
N\"{o}thnitzer Str. 40, 01187 Dresden, Germany}

\author{F. Steglich}
\affiliation{Max Planck Institute for Chemical Physics of Solids, N\"{o}thnitzer Str. 40,
01187 Dresden, Germany}



\date{\today}
\begin{abstract}
We studied the temperature-pressure phase diagram of
EuFe$_{2}$As$_{2}$ by measurements of the electrical resistivity.
The antiferromagnetic spin-density-wave transition at $T_0$
associated with the FeAs-layers is continuously suppressed with
increasing pressure, while the antiferromagnetic ordering
temperature of the Eu$^{2+}$ moments seems to be nearly pressure
independent up to 2.6 GPa. Above 2 GPa a sharp drop of the resistivity, $\rho(T)$, indicates the onset of superconductivity at $T_c\approx29.5$~K. Surprisingly, on further reducing the temperature $\rho(T)$ is increasing again and exhibiting a maximum caused by the ordering of the Eu$^{2+}$ moments,
a behavior which is reminiscent of {\it re-entrant} superconductivity as it is observed in the  ternary Chevrel phases or in the rare-earth nickel borocarbides.

\end{abstract}


\maketitle


The recent discovery of high temperature superconductivity (SC) in the \textit{R}FeOAs
compounds (\textit{R}=La-Gd) \cite{kamihara, xchen, cruz, clauss, fratini} with
superconducting transition temperature approaching values up to 50~K \cite{kamihara,
xchen, fratini} attracted strong interest in the scientific community. The
\textit{R}FeOAs compounds forming in the ZrCuSiAs-type tetragonal structure are closely
related to \textit{A}Fe$_2$As$_2$ (\textit{A}=Ca, Sr, Ba) \cite{ni,krellner,rotter1}
forming in the ThCr$_2$Si$_2$-type tetragonal structure. Both share a similar arrangement
of Fe$_2$As$_2$ layers assumed to be the key for the SC formation in this class of
compounds. Materials from both families show at $T_0\approx 150{\rm~ K}-210$~K  a
structural transition from a tetragonal to an orthorhombic phase  which is closely
related to the formation of a spin-density-wave (SDW) type magnetic instability
\cite{ni,krellner, sasmal, gchen1, jeevan}.

In contrast to the \textit{A}Fe$_2$As$_2$ (\textit{A}=Ca, Sr, Ba) compounds where only
the iron possesses a magnetic moment, in EuFe$_2$As$_2$ a large additional magnetic
moment of $7\mu_B$ is carried by Eu which is in the $2+$ state. Like the (\textit{A}=Ca,
Sr, Ba) members of the \textit{A}Fe$_2$As$_2$ family, EuFe$_2$As$_2$ exhibits a SDW
transition around $T_0=190$~K related to the Fe$_2$As$_2$ layers, but additionally at
$T_N=20$~K the magnetic moments of the localized Eu$^{2+}$ moments order
\cite{raffius,jeevan,ren} in a so called A-type antiferromagnetic structure \cite{jiang}.
SrFe$_2$As$_2$ has similar structural properties, the unit-cell volume is only 3\%
larger, and a comparable value of $T_0=210$~K \cite{krellner,jesche}. Furthermore, aside
from the Eu 4{\it f} part, the electronic density of states (DOS) of EuFe$_2$As$_2$ is
almost identical to the DOS of SrFe$_2$As$_2$ \cite{jeevan}. Therefore, SrFe$_2$As$_2$
can be considered as a non-magnetic homologue compound of EuFe$_2$As$_2$. Provided that
the two different kinds of magnetic ordering phenomena are reasonably well decoupled the
results of previous doping and pressure studies on SrFe$_2$As$_2$
\cite{alj,alireza,kumar} would suggest the appearance of a superconducting phase on
doping and/or at high pressure in EuFe$_2$As$_2$ too. Indeed, a very recent K doping
investigation \cite{jeevan2} confirmed the first prediction:
K$_{0.5}$Eu$_{0.5}$Fe$_2$As$_2$ is superconducting below $T_c=30$~K. As a result of the
replacement of half of the Eu by K, no clear signature of the ordering of the Eu$^{2+}$
is present anymore.

Our electrical resistivity measurements under hydrostatic pressure on single crystalline EuFe$_2$As$_2$
 indicate that the SDW transition is continuously suppressed upon applying
pressure, while the magnetic ordering of the Eu$^{2+}$ moments seems to be very robust
against pressure. Above 2~GPa a sharp drop of the resistivity indicates the emergence of
a superconducting phase below $T_c\approx29.5$~K  value which is close to the one
reported for $T_c$ for K$_{0.5}$Eu$_{0.5}$Fe$_2$As$_2$ \cite{jeevan2}. For the first time
to our knowledge in FeAs based superconductors, we found an indication of {\it
re-entrant} SC as observed in the ternary Chevrel phases, e.g. GdMo$_6$S$_8$
\cite{ishikawa} or in the rare-earth nickel borocarbides, e.g. HoNi$_2$B$_2$C
\cite{eisaki}.

Single crystals of EuFe$_2$As$_2$ were synthesized using the
Bridgman method \cite{jeevan}. Powder X-ray diffraction confirmed
the proper ThCr$_2$Si$_2$ type tetragonal structure and the single
phase nature of the sample. Measurements of the electrical
resistance were carried out using a standard four-probe technique
with current flowing in the $(a,b)$-plane and magnetic field applied
parallel to the current. The investigations were done from room temperatures down to 1.8 K and in magnetic
fields up to 7~T using a physical property
measurement system (PPMS, Quantum Design). Pressures up to 2.6 GPa
have been generated using a double-layer piston-cylinder type
pressure cell with an inner cylinder made from MP35N. Silicone oil
was used as pressure transmitting medium. The superconducting
transition temperature of lead served as a pressure gauge. The
narrow transition width confirmed the quasi hydrostatic pressure
conditions inside the pressure cell. The density functional band
structure calculations within the local (spin) density approximation
(L(S)DA) have been performed using a full potential code FPLO
\cite{fplo}. The strong Coulomb repulsion in the Eu 4$f$ orbitals
have been included in a mean field level using the atomic limit
double counting scheme \cite{czyzyk} in the LSDA+$U$ approximation.
A $U$ value of 8 eV was used for the Eu 4$f$ orbitals, but the
resultant conclusions did not change for a range from 6 to 10 eV.
The total energies were calculated on a dense mesh of $k$-points
using the Perdew-Wang \cite{perdew} exchange correlation potential.

\begin{figure}[t]
\centering
\includegraphics[angle=0,height=7cm,clip]{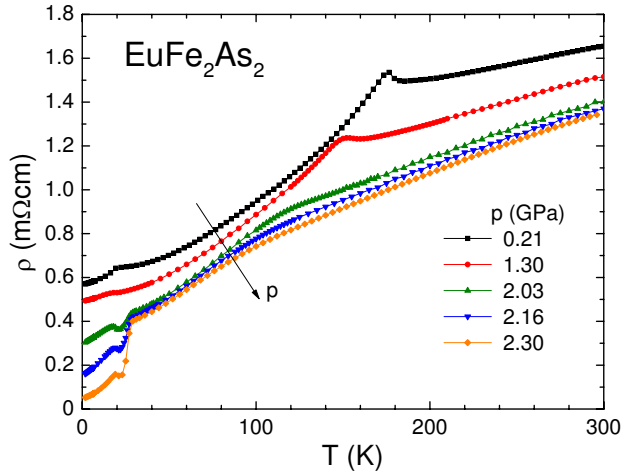}
\caption{\label{rhoT} (Color online) Electrical resistivity {\it
vs.} temperature for different applied pressures for single
crystalline EuFe$_2$As$_2$.}
\end{figure}

\begin{figure}[t!]
\includegraphics[height=6.5cm ,angle=0]{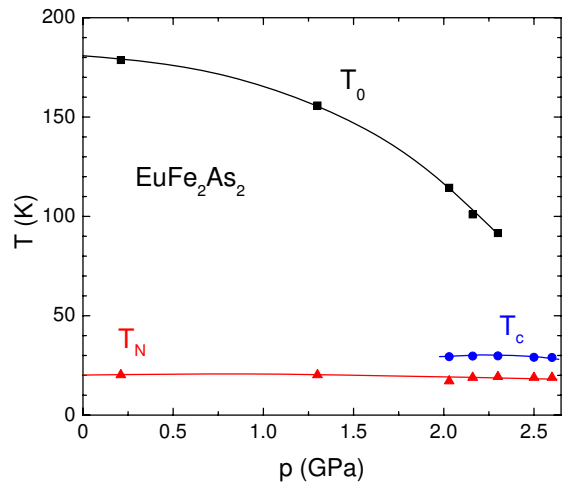}
\caption{\label{phasediagram} (Color online) Pressure-temperature
phase diagram of EuFe$_2$As$_2$ obtained from $\rho(T)$ data. The
lines are guides to the eye.}
\end{figure}

Figure \ref{rhoT} shows the electrical resistivity, $\rho(T)$ of EuFe$_2$As$_2$ for
different applied pressures. The absolute values of the room temperature resistivity
$\rho(300K)\approx1.6$ m$\Omega$cm at ambient pressure is typical for the
\textit{A}Fe$_2$As$_2$ materials. In the whole investigated pressure range the
resistivity decreases continuously on decreasing temperature, with the exception of a
clear anomaly indicating SDW type of magnetic transition at $T_0$ at low pressures. With
increasing pressure the peak broadens. At $p=2.3$~GPa only a change of slope in $\rho(T)$
is remaining, however, $T_0$ can be still determined from the minimum in the second
derivative of the resistivity, ${\rm d^2}\rho(T)/{\rm d}T^2$. At higher pressures this
anomaly cannot be any longer unanimously detected. At low temperature a second anomaly
appears around $T_N\approx20$~K indicating the magnetic ordering of the Eu$^{2+}$ moment.
$T_N(p)$ seems to be nearly pressure independent. At $p=2.03$~GPa for the first time a
sharp drop of the resistivity appears around $T_c=29.5$~K. With increasing pressure this
drop becomes even sharper and more pronounced, while its position does not change. We
ascribe this feature in the resistivity to the onset of SC. The complete formation of the
superconducting state is interrupted by the ordering of the Eu$^{2+}$ sublattice at
$T_N<T_c$. The ordering of the Eu$^{2+}$ causes an initial increase of $\rho(T)$ followed
by a maximum on lowering the temperature. This behavior of $\rho(T)$ is reminiscent of
{\it re-entrant} SC which has first been reported in GdMo$_6$S$_8$ \cite{ishikawa} which
belongs to the class of the ternary Chevrel phases. Furthermore, {\it re-entrant} SC has
been found in the rare-earth nickel borocarbides (e.g. HoNi$_2$B$_2$C) \cite{eisaki}. In
this pressure range ($p\geq2.16$) we take  $T_N$ as the temperature where the resistivity
reaches its maximum below $T_c$. In the rare-earth nickel borocarbides this has been
shown to be the proper procedure to determine $T_N$ from resistivity in this region of
the phase diagram. The resulting pressure-temperature phase diagram is presented in
Fig.~\ref{phasediagram}.

\begin{figure}[t]
\includegraphics[height=7cm,angle=0]{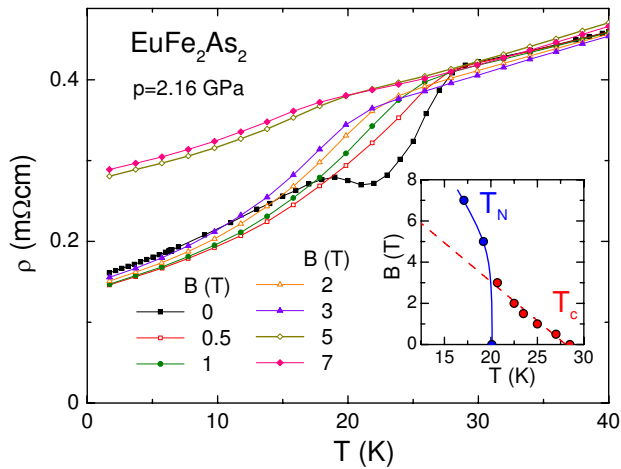}
\caption{\label{rhoTH} (Color online) Electrical resistivity at
$p=2.16$~GPa in different magnetic fields. Inset:
temperature-magnetic field phase diagram. See text for details.}
\end{figure}

To get further insights in the relation of magnetic ordering of the Eu$^{2+}$ moments and
SC in EuFe$_2$As$_2$ under pressure we measured $\rho(T)$ at $p=2.16$~GPa in different
external magnetic fields (cf. Fig.~\ref{rhoTH}). As discussed before the zero field
resistivity data show the typical behavior found in re-entrant superconductors: after a
first drop at $T_c$, $\rho(T)$ exhibits a maximum at $T_N$ before it further decreases
upon lowering the temperature. On increasing magnetic field the drop in resistivity
shifts to lower temperatures. Already at a field of $B=0.5$~T no maximum in $\rho(T)$ is
visible anymore. Although a drop in resistivity is present in the whole investigated
field range up to $B=7$~T there is a qualitative difference between magnetic fields
$B\leq3$~T and $B\geq5$~T. In the first case $\rho(T)$ is decreasing much stronger
compared with the latter, dividing the data sets in two distinct groups. The small
reduction of the resistivity in large magnetic fields is similar in size to what is found
at lower pressures, where no SC is present, below $T_N$ due to the reduced scattering of
the conduction electrons on the ordered Eu$^{2+}$ moments. To construct a $T-B$ phase
diagram we take the temperature $T_x$ of minima of the second derivative of $\rho$, ${\rm
d^2}\rho(T)/{\rm d}T^2$, which is corresponding to the kink in $\rho(T)$. The phase
diagram is depicted in the inset of Fig.~\ref{rhoTH}. At low fields $B_x(T)=B(T=T_x)$ is
increasing linearly. Between 3~T and 5~T, $B_x(T)$ shows a strong upturn. Having in mind
the smaller drop of $\rho(T)$ in large magnetic fields we suggest that the kink in
resistivity at $B=5$~T and 7~T should be attributed to the magnetic ordering of the
Eu$^{2+}$ moments. Therefore we include the data point obtained for $T_N$ at zero
magnetic field to propose a magnetic phase line. The resulting $T-B$ phase diagram
suggests the suppression of the SC phase once the $H_{c2}$ line crosses the magnetic
phase boundary. We find an initial slope of $\partial B_{c2}(T)\partial T|_{T_c}=-0.368
{\rm~ T/K}$ much smaller compared to the value found in SrFe$_2$As$_2$ ($\partial
B_{c2}(T)/\partial T|_{T_c}=-2.05 {\rm~ T/K}$) \cite{kumar}. Accordingly, the estimated
orbital critical field for superconductivity is only about $B_{c2}(0)\sim19.5$ T.

A recent study of magnetization and magnetoresistance at atmospheric pressure reports a
metamagnetic transition to a ferromagnetically polarized state in EuFe$_2$As$_2$ already
at about $B_m=1$~T at $T=2$~K for field applied in the $ab$-plane. Under pressure $B_m$
seems to be much higher. Our resistivity measurements at $p=2.16$~GPa point to the
presence of AFM order in magnetic fields as high as 7~T. A possible explanation for the
more robust AFM state could be the higher compressibility along $c$ axis compared with
$a$ axis as has been found in SrFe$_2$As$_2$ \cite{deepa}. This should lead to a stronger
magnetic coupling along the $c$ axis upon increasing pressure.

Recently it has been shown that there is a volume collapse under pressure that precedes
the onset of superconductivity in CaFe$_{2}$As$_{2}$\cite{Kreyssig,yildirim}. Thus the
first problem we addressed with our band structure calculations as a function of pressure
for EuFe$_{2}$As$_{2}$ was the possibility of a similar volume collapse behavior and the
ramifications to the SDW of Fe (suppressed or not) and the Eu valency (possible valence
transition from Eu$^{2+}$ to Eu$^{3+}$). The internal parameter for the As $z$ position
was held fixed at the experimental position of the ambient pressure during these
calculations, while the $c/a$ ratio was optimized at different reduced volumes. The
calculations were done with the Fe spin-moments aligned in the columnar magnetic
structure pattern, as observed for the Sr analogue \cite{jesche}. Our results do not
indicate any tendency to a sudden collapse of the $c$ axis under pressure. The SDW is
fully suppressed for pressures larger than 5 GPa, while the Eu is stable in the 2+ state
up to pressures larger than 10 GPa.

Another interesting question to answer is the strength of the antiferromagnetic
interaction between the Eu$^{2+}$ planes as a function of pressure. To evaluate this, we
have calculated the energy difference between the ferromagnetic and anti-ferromagnetic
(inter planes) arrangement of the Eu spins for various pressure values \cite{footnote}.
First results from the LSDA+$U$ calculations show that pressure stabilizes the AF
ordering of the Eu$^{2+}$ planes along the $c$-axis. The influence of spin-orbit coupling
in this scenario is still under investigation. As a next step we have compared the change
in the magnitude of the Fe moments as a function of pressure for both the Sr and Eu
systems. The magnitude of the Fe moments, ordered in the columnar magnetic structure
decreases upon reducing volume and finally is suppressed in a similar trend for both
systems.

Comparing the electronic structure of EuFe$_{2}$As$_{2}$ to the Sr analogue, a very close
similarity remains, under pressure, between these two systems. The $T_{c}$ for the 50\%
K-substituted EuFe$_{2}$As$_{2}$ is reduced by 6 K as compared to the 50\% K-substituted
SrFe$_{2}$As$_{2}$. Also, in the current pressure study of the pure Eu system, a similar
reduction (by 6 K) of the $T_{c}$ is observed compared to the Sr analogue. But both the
Sr and Eu analogues behave very similar at high temperatures (above the Eu magnetic
ordering temperature). Therefore one can conclude that this decrease of the $T_{c}$
together with the reduction of the $\partial B_{c2}/\partial T |_{T_{c}}$ slope are
caused by the presence of paramagnetic Eu$^{2+}$ ions at temperatures larger than the
N\'{e}el ordering temperature.

In summary, we have investigated the effect of hydrostatic pressure on the peculiar
properties of EuFe$_2$As$_2$. The transition at $T_0$ corresponding to the lattice
distortion and the formation of the SDW shifts to lower temperatures with increasing
pressure, down to 90 K at $p=2.3$ GPa. The corresponding anomaly in $\rho(T)$ has already
become very weak at this pressure and cannot be any longer observed for $p\geq 2.5$~GPa,
suggesting the critical point were this transition gets completely suppressed to be very
close to $p=2.3$~GPa. Above 2 GPa, a sharp drop appears in $\rho(T)$ at $T_c = 29.5$~K
and becomes more pronounced with further increasing pressure, indicating the onset of
superconductivity. The large and linear initial shift of $T_c$ to lower temperatures with
increasing field supports the SC nature of this transition. Thus the transition from the
magnetic order of Fe moments to the SC state occurs in EuFe$_2$As$_2$ at a slightly
smaller pressure than in SrFe$_2$As$_2$, which correlates with a slightly smaller initial
unit cell volume. Further on, assuming that the transition at $T_0$ is first order (as
suggested by most current investigations) and thus the applicability of the
Clausius-Clapeyron relation, the faster suppression of $T_0$ with increasing $p$ is in
accordance with a larger volume change at $T_0$ in EuFe$_2$As$_2$ ($\Delta V \approx -5
\times 10^{-4} {\rm nm}^3$) compared to SrFe$_2$As$_2$ ($\Delta V \approx -3 \times
10^{-4} {\rm nm}^3$) \cite{Tegel new}, while the latent heat at the transition, $\Delta H
\approx 220$ J/mol \cite{jeevan}, is about the same as in SrFe$_2$As$_2$ \cite{krellner}.
In the SC state, we observed below $T_c$ a minimum in $\rho(T)$ followed by a clear
increase leading to a maximum at around 20 K. This suggests re-entrant superconductivity
due to the antiferromagnetic ordering of Eu. The Eu antiferromagnetic ordering
temperature $T_N$ itself does not seem to change significantly with pressure, as
typically observed in Eu systems. While this re-entrant behavior is suppressed at low
fields $B\geq 0.5$~T, we observed between $B = 3$~T and $B = 5$~T a strong reduction of
the drop in $\rho(T)$ below the transition as well as a strong reduction of the field
dependance of the transition temperature. This suggests that once $T_c$ becomes smaller
than $T_N$, superconductivity is completely suppressed and the left anomaly in $\rho(T)$
observed at $T_x$ is then related to $T_N$. Thus our results evidence a very peculiar and
very interesting interaction between the superconducting state and the magnetism of the
rare earth in EuFe$_2$As$_2$ under pressure. Such a re-entrant superconductivity as well
as the suspected suppression of the SC state by the antiferromagnetic state at higher
field have not been observed in the doped RFeAsO compounds and makes thus EuFe$_2$As$_2$
unique among the layered FeAs systems. The occurrence of these phenomena seems to be
related to the fact that at $B = 0$, $T_N$ is not much smaller than $T_c$. That the
magnetic order of Eu is able to suppress the superconductivity in the FeAs layers has
likely strong implication for the symmetry of the SC order parameter.

We acknowledge the financial support of the DFG - MI 1171/1-1, DFG - Research Unit 960
and BRNS (grant no. 2007/37/28).


\begin{thebibliography}{ }
\bibitem{kamihara}Y. Kamihara, T. Watanabe, M. Hirano, and H. Hosono, J. Am. Chem. Soc. {\bf130}, 3296 (2008).

\bibitem{xchen}X. F. Chen, T. Wu, G. Wu, R. H. Liu, H. Chen, and D. F. Fang, Nature {\bf453}, 761 (2008).

\bibitem{cruz}C. de La Cruz, Q. Huang, J. W. Lynn, Jiying Li, W. Ratcliff II, J. L. Zarestky, H. A. Mook, G. F. Chen,
J. L. Luo, N. L. Wang, and P. Dai, Nature {\bf453}, 899 (2008).

\bibitem{clauss}H. H. Clauss, H. Luetkens, R. Klingeler, C. Hess, F.J. Litterst, M. Kraken, M. M. Korshunov,
I. Eremin, S. L. Drechsler, R. Khasanov, A. Amato, J. Hamann-Borreo,
N. Leps, A. Kondrat, G. Behr, J. Werner, and B. B\"{u}chner,
arXiv:0805.0264 (unpublished).

\bibitem{fratini}M. Fratini, R. Caivano, A. Puri, A. Ricci,
Z. A. Ren, X. L. Dong, J. Yang, W. Lu, Z. X. Zhao, L. Barba, G.
Arrighetti, M. Polentarutti, and A. Bianconi, Supercond. Sci.
Techno. {\bf21}, 092002 (2008).

\bibitem{raffius}H. Raffius, E. M\"{o}rsen, B. D. Mosel, W.
M\"{u}ller-Warmuth, W. Jeitschko, L. Terbb\"{u}chte, and T. Vomhof,
J. Phys. Chem. Solids {\bf54}, 135 (1993).

\bibitem{ni}N. Ni, S. Nandi, A. Kreyssig, A. I. Goldman, E. D. Mun, S. L.
Bud'ko, P. C. Canfield, arXiv:0806.4328 (unpublished).

\bibitem{krellner}C. Krellner, N. Caroca-Canales, A. Jesche, H. Rosner, A. Ormeci, and C. Geibel, arXiv:0806.1043 (unpublished).

\bibitem{rotter1}M. Rotter, M. Tegel, D. Johrendt, I. Schellenberg, W. Hermes, and R. P\"{o}ttgen,
Phys. Rev. B {\bf78}, 020503(R) (2008).

\bibitem{jeevan}H. S. Jeevan, Z. Hossain, D. Kasinathan, H. Rosner, C. Geibel, P.
Gegenwart, Phys. Rev. B {\bf78}, 052502 (2008).

\bibitem{sasmal}K. Sasmal, B. Lv, B. Lorenz, A. M. Guloy, F. Chen, Y. Y.
Xue, and C. W. Chu, arXiv:0806.1301 (unpublished).

\bibitem{gchen1}G. F. Chen, Z. Li, J. Dong, G. Li, W. Z. Hu, X. D.
Zhang, X. H. Song, P. Zheng, N. L. Wang, and J. L. Luo,
arXiv:0806.2648 (unpublished).

\bibitem{ren}Z. Ren, Z. Zhu, S. Jiang, X. Xu, Q. Tao, C. Wang, C. Feng, G. Cao, and
Z. Xu, Phys. Rev. B {\bf78}, 052501 (2008).

\bibitem{jiang}S. Jiang, Y. Luo, Z. Ren, Z. Zhu, C. Wang,
X. Xu, Q. Tao, G. Cao, and Z. Xu, arXiv:0808.0325 (unpublished).

\bibitem{jesche}A. Jesche, N. Caroca-Canales, H. Rosner, H. Borrmann,
A. Ormeci, D. Kasinathan, K. Kaneko, H. H. Klauss, H. Luetkens, R.
Khasanov, A. Amato, A. Hoser, C. Krellner, and C. Geibel,
arXiv:0807.0632 (unpublished).

\bibitem{kumar}M. Kumar, M. Nicklas,
A. Jesche, N. Caroca-Canales, M. Schmitt, M. Hanfland, D.
Kasinathan, U. Schwarz, H. Rosner, C. Geibel, arXiv:0807.4283
(unpublished).

\bibitem{alireza}P. L. Alireza, J. Gillett, Y. T. C. Ko, S. E. Sebastian, and G. G.
Lonzarich, arXiv:0807.1896 (unpublished).

\bibitem{alj}A. L. Jasper, W. Schnelle, C. Geibel, and H. Rosner, arXiv:0807.2223 (unpublished).

\bibitem{jeevan2} H. S. Jeevan, Z. Hossain, Deepa Kasinathan, Helge Rosner, C. Geibel, and P. Gegenwart, arXiv:0807.2530 (unpublished).

\bibitem{ishikawa}M. Ishikawa and {\O}. Fischer, Solid State
Comm. {\bf23}, 37 (1977).

\bibitem{eisaki}H. Eisaki, H. Takagi, R. J. Cava, B. Batlogg, J. J. Krajewski, W. F. Peck,
Jr., K. Mizuhashi, J. O. Lee, and S. Uchida, Phys. Rev. B {\bf50},
647 (1994).

\bibitem{fplo} K. Koepernik, and H. Eschrig, Phys. Rev. B {\bf 59}, 1743
(1999); I. Ophale, K. Koepernik, and H. Eschrig, Phys. Rev. B {\bf
60}, 14035 (1999); http://www.fplo.de.

\bibitem{czyzyk}M. T. Czy\.{z}yk and G. A. Sawatzky, Phys. Rev. B {\bf 49},
14211 (1994).

\bibitem{perdew} J. P. Perdew and Y. Wang, Phys. Rev. {\bf B 45}, 13244
  (1992).







\bibitem{deepa} D. Kasinathan (private communication).


\bibitem{Kreyssig}A. Kreyssig, M. A. Green, Y. Lee, G. D. Samolyuk, P. Zajdel, J. W. Lynn,
S. L. Bud'ko, M. S. Torikachvili, N. Ni, S. Nandi, J. Le\~{a}o, S.
J. Poulton, D. N. Argyriou, B. N. Harmon, P. C. Canfield, R. J.
McQueeney, and A. I. Goldman, arXiv:0807.3032 (unpublished).


\bibitem{yildirim} T. Yildirim, arXiv:0807.3936 (unpublished).

\bibitem{footnote} To describe the magnetic interactions as
realistic as possible, the lattice parameters for these calculations
for EuFe$_{2}$As$_{2}$ have been scaled from the X-ray diffraction
data under pressure for the non-magnetic homologue
SrFe$_{2}$As$_{2}$ \cite{kumar}.

\bibitem{Tegel new} M. Tegel, M. Rotter, V Weiss, F. M. Schappacher, R. Poettgen, and D. Johrendt, arXiv:0806.4782 (unpublished).


\end{thebibliography}
\end{document}